# KEYWORD SPOTTING SYSTEM AND EVALUATION OF PRUNING AND QUANTIZATION METHODS ON LOW-POWER EDGE MICROCONTROLLERS

*Jingyi Wang, Shengchen Li*

Xi'an Jiaotong-Liverpool University
School of Advanced Technology, No.111 Ren'ai Road
Suzhou, Jiangsu 215123, China
Wangjingyi70@gmail.com, Shengchen.Li@xjtlu.edu.cn

**ABSTRACT**

Keyword spotting (KWS) is beneficial for voice-based user interactions with low-power devices at the edge. The edge devices are usually always-on, so edge computing brings bandwidth savings and privacy protection. The devices typically have limited memory spaces, computational performances, power and costs, for example, Cortex-M based microcontrollers. The challenge is to meet the high computation and low-latency requirements of deep learning on these devices.

This paper firstly shows our small-footprint KWS system running on STM32F7 microcontroller with Cortex-M7 core @216MHz and 512KB static RAM. Our selected convolutional neural network (CNN) architecture has simplified number of operations for KWS to meet the constraint of edge devices. Our baseline system generates classification results for each 37ms including real-time audio feature extraction part.

This paper further evaluates the actual performance for different pruning and quantization methods on microcontroller, including different granularity of sparsity, skipping zero weights, weight-prioritized loop order, and SIMD instruction. The result shows that for microcontrollers, there are considerable challenges for accelerate unstructured pruned models, and the structured pruning is more friendly than unstructured pruning. The result also verified that the performance improvement for quantization and SIMD instruction.

*Index Terms*— Keyword spotting, embedded deep learning, microcontroller, pruning, quantization

## 1. INTRODUCTION

Keyword spotting (KWS) helps human interact with computers by voice, which is useful when one is inconvenient to use a keyboard and mouse. KWS may be defined as a task that identifies keywords from a sound containing speech [1]. Recently, such an AI application is commonly running on the edge, such as smart-phone, microcontroller, IoT device for various scenarios. KWS may be used to detect some command words like "left", "right", "yes", or wake-up words such as "Hey Siri", "Ok Google", etc. [2]. The edge device is typically always-on to identify these words, and may be activated when wake-up words detected, then higher cost speech recognition model may run on the edge or in the cloud. Such an edge computing workflow brings power savings, bandwidth savings, privacy protection, reliability, low-cost and low-latency [3], [4]. A general pipeline of KWS implementation consists of audio feature extraction, neural network (NN), and posterior handing [1].

However, there are some limits and challenges for deploying NNs on microcontrollers [3], take STM32 as example:

- **Limited Memory Footprint:** The random-access memory (RAM) is used to storage input/output, weights, activation data at running. The size of on-chip RAM (SRAM) is various from 20KB to 512KB. An external RAM (DRAM) can be added to the board with larger size, higher power and lower speed. The read-only memory (FLASH) is used to storage weights when power-off. The size of on-chip FLASH is usually between 64KB and 1MB.
- **Limited Computing Speed:** The number of operations per second is limited. The CPU (Cortex-M) frequency is typically between 72Mhz and 216MHz.

These limits from hardware limit the NN models in two ways: the number of parameters and the number of operations of the model. To address the limits, the corresponding deep neural networks (DNNs) [5], convolutional neural networks (CNNs) [6], convolutional recurrent neural networks (CRNNs) [7] and long-short term memory (LSTM) networks [8] for small-footprint KWS were proposed. Since there are many redundant parameters in NNs, pruning method for NNs was proposed by Han et. al [9]. The work shows that about 90% of weights may be removed for convolutional layers without loss of accuracy. Consider unstructured pruning may be hard to be accelerate, more regular pruning methods with various granularities were proposed and discussed [10], also known as structured pruning. The trade-off is between regularity and accuracy. In addition, quantitative methods for KWS [11] and other NN models have been proposed and widely used.

In Section 2, this paper firstly shows our small-footprint KWS system as our baseline system running on STM32F7 microcontroller with ARM Cortex-M7 core @216MHz and 512KB static RAM. Our selected CNN architecture has simplified number of operations and parameters for KWS to meet the limitation of microcontrollers. There is only one convolutional layer and no pooling layers in the architecture. In Section 3, this paper further evaluates the impacts of different pruning, quantification methods on model inference time and power consumption in our real system. We focus on the real performance of pruning with various granularities and various quantitative data types, and try to accelerate the unstructured pruned NNs by adding conditional statements and changing order of nested loops of convolutional layers. In addition, we compared the performance with/without Single Instruction/Multiple Data (SIMD) instruction of Cortex-M microcontrollers [12], which allows multiple data to be processed in a single instruction.

Our code for the experiments is available at:
https://github.com/RoboBachelor/Keyword-Spotting-STM32

## 2. SMALL FOOT-PRINT KEYWORD SPOTTING SYSTEM ON MICROCONTROLLER

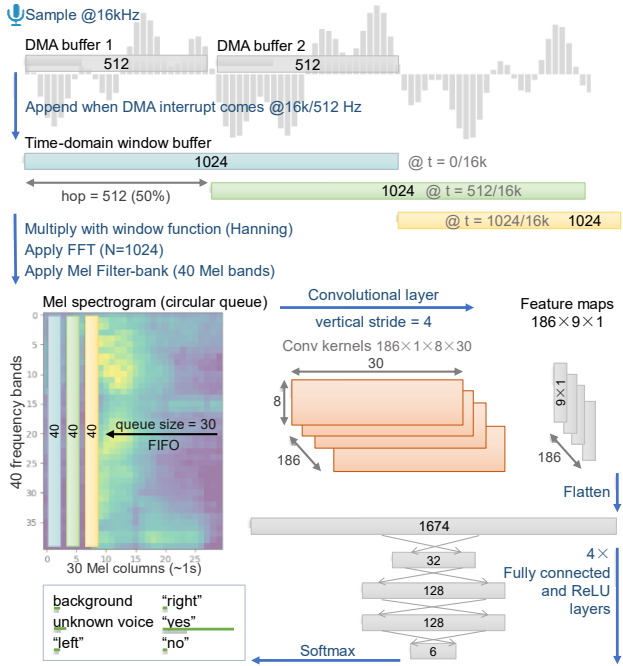

**Figure 1** The overall structure of our baseline smart foot-print keyword spotting system on STM32F7 microcontroller.

### 2.1. Model Architecture and Training

The overall structure of our baseline small foot-print KWS on STM32F7 microcontroller is shown in Figure 1. Our CNN model architecture is almost the same as *"cnn-one-fstride4"* described in this work [6]. The convolutional kernel width is 30, same as the width in time axis of input spectrogram, so there is no stride and pooling in time axis. The kernel height in frequency axis is 8 and the vertical stride is 4 (50% overlap in frequency). The pooling (down sampling) in frequency is no longer needed and substituted by vertical stride. After a single convolutional layer, the output feature maps are directly flattened and feed into a 4-layer-DNN. Finally, the 6 output neurons are softmax proceeded and mean the 6 target labels. Speech Command Data Set v0.01 [13] is used and 4 keywords are selected: "yes", "no", "left" and "right". A "background" category is added by using sound data in background folder of dataset, and an extra "unknown voice" is added by sampling the voices in other categories. There are 2000 sound clips are selected for training and other 350 clips are for testing. After 20 epochs of training, the validation accuracy is about 90%.

### 2.2. Feature Extraction and Deployment on Microcontroller

Our development board is Apollo STM32F767 Development Board by Alientek [14]. The whole hardware experiment platform is shown in Figure 2. The microcontroller STM32F767IGT6 is based on Cortex-M7 @ 216MHz. The used resources on chip and board are labelled on the figure. A real-time operating system, FreeRTOS, runs on the board, used to schedule the feature extraction task and model inference task, and response to the interruptions. The data (except DMA buffers) is stored as 32-bit float number, and floating-point unit (FPU) of STM32 [15] is used to accelerate float operations. All of the codes are written in C and compiled with fastest option by ARM compiler V6. A hardware timer (TIM) is used to obtain elapsed time.

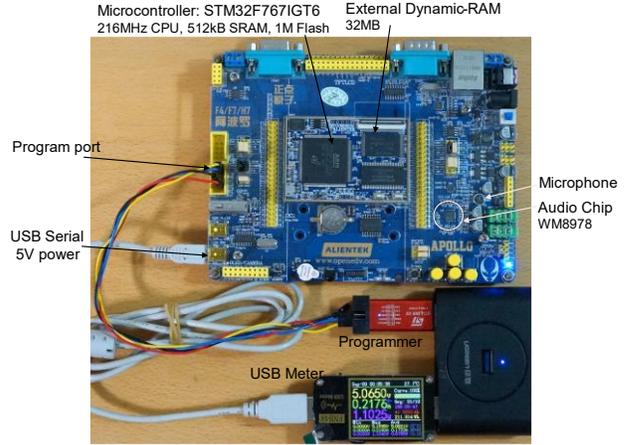

**Figure 2** Apollo STM32F767 development board and experiment configuration

The audio signal is sampled at 16kHz frequency, stereo and copied from WM8978 chip to Direct Memory Access (DMA) double buffers on microcontroller via I2S protocol automatically after initialize. Once DMA interruption comes when one buffer is full for each ~31ms, one channel of the new-coming sound is normalized to float and copied to the rear of time-domain window buffer in Figure 1, and the previous sound data is moved to the front of this buffer. In other words, at each DMA interruption, 50% old sound data is shifted-out to left and 50% new sound data is shifted-in from right. The window buffer is firstly multiplied by Hanning window function sample-wisely, secondly applied 1024-point FFT, thirdly applied 40-band Mel filter-bank by multiplying the pre-calculated Mel coefficient table stored in microcontroller, and finally convert to log amplitude, to generate a Mel column. The Mel spectrogram is a circular queue, treats each Mel column as an element and stores the latest 30 columns (duration ~1s) as the Mel spectrogram. The whole feature extraction process costs about 6ms.

The model inference task runs loopily and has lower priority than feature extraction task. The feature extract task not stops and may interrupt the inferencing process when model is inferencing to achieve the real-time property. The parameters are stored in on-chip SRAM, while the two activation data buffers (with largest possible sizes, 6.5KB and 0.5KB) are stored in DRAM. The total number of parameters is 119,936 (44,826 for conv layer, 75,110 for 4 FC layers) and takes 468.5KB memory. The storage order of convolutional kernels in memory is *out-channel → in-channel → row → col*. The computation loop order for convolutional layer is *out-channel → out-row → out-col → in-channel → kernel-row → kernel-col*. The total number of multiply–accumulates (MACs) is 476,576 (401,760 for conv layer, 74,816 for FC layers). At the beginning of inference, the latest spectrogram is obtained from feature extraction task (with FreeRTOS mutex) and normalized to normal distribution. It takes about 31ms for single inference.

## 3. EVALUATION OF PRUNING AND QUANTIZATION METHODS

### 3.1. Various Granularities of Pruning and Sparsity

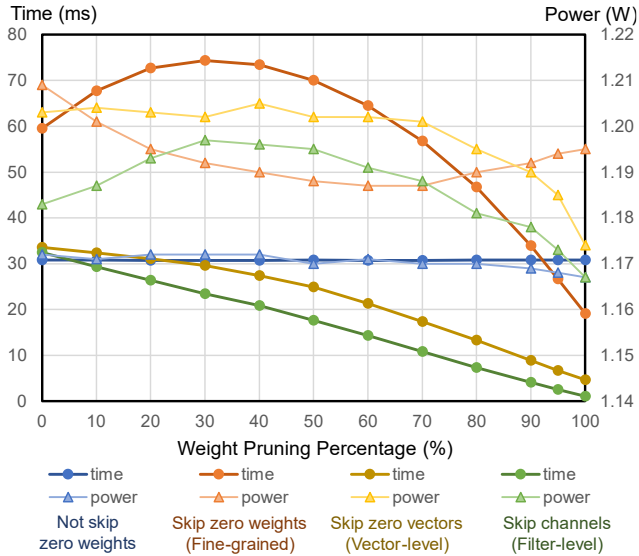

**Figure 3** Time and power consumption with various granularities and percentages of pruning

According to this work [10], there are 4 different structures of sparsity in a 4-dimensional convolutional weight array: fine-grained, vector-level, kernel-level and filter-level sparsity. Fine-grained sparsity is also known as unstructured pruning that removes individual weights and has higher accuracy. Vector-level, kernel-level and filter level are structured pruning, that correspondingly removes rows or columns in kernels, whole kernels, and groups of kernels belong to same channels. Since our model has only one input channels, the kernel-level and filter-level sparsity are equivalent to reduce number of output channels of the model architecture. The consumed inference time and power with various sparsity are shown in Figure 3. In the second experiment configuration, an "if" statement is added to compare the weight value and skip the floating-point calculations for zero weights. For the vector-level configuration, a 3-dimensional mask array is used to represent removed rows of kernels and an "if" statement is also added. For the filter-level, the number of channels was reduced in the code. The results show that:

- Without "if" statement, the computation time and power are hardly affected by how many weight values are zero;
- For fine-grained sparsity, an additional "if" statement may bring negative effects of time since the pipeline of instruction execution may be interrupted. Also, the compiler may not understand code in the way we expect;
- For structured pruning in vector-level sparsity and filter-level sparsity, the advantages start to appear when small percentage of weights are moved.

### 3.2. Weight Prioritized Loop for Unstructured Pruning

Since the advantage of unstructured pruning is quite small in previous results, we try to further accelerate the unstructured pruned convolutional layers by changing the nested loop order. The common loop order is to iterate each value in output feature maps and load the same kernel value for multiple times. Our weight prioritized loop order iterates the kernel values, and complete all the computations related to an individual kernel value. Ignoring the output channel and input channel dimensions, the two loop orders are illustrated and compared in Figure 4, and the numbers of memory access/storage times and numbers of "if" statement execution times are also given. For weight prioritized order, if a weight value is zero, all its relevant computing is skipped by just using one "if" statement.

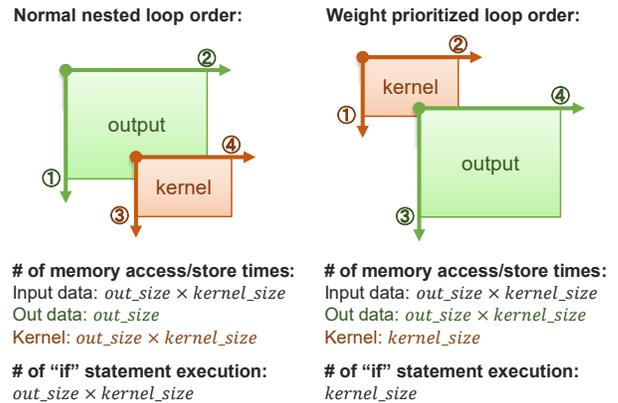

**Figure 4** Illustration of normal and weight prioritized nested loop orders for computational layer

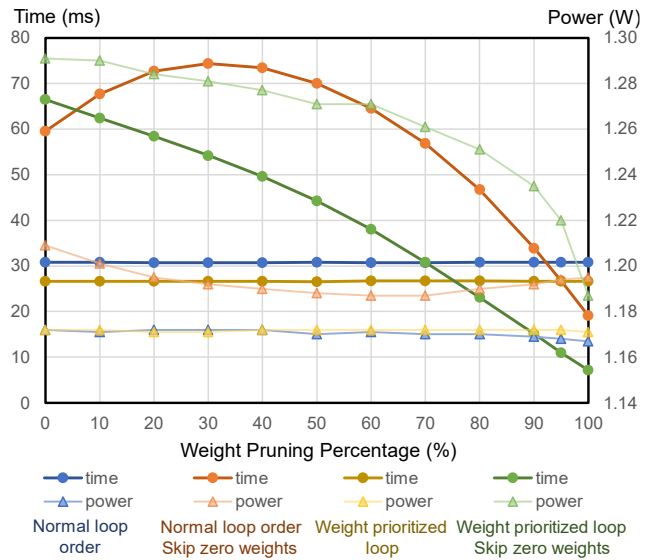

**Figure 5** Time and power consumption for normal and weight prioritized loop orders with/without skipping zero weights

The time and power consumption for these two orders with/without skipping zero weights is shown in Figure 5. Without "if" statement to skip zero weights, the weight prioritized loop is a bit faster. With weight prioritized loop, the unstructured pruned

CNN is faster when pruning percentage larger than 80% and the speed reaches twice when 90% of weights removed. However, the weight prioritized loop combined with skipping zero weights may be incompatible with other accelerating methods for microcontrollers [3].

### 3.3. Quantization and SIMD Instruction

Representing weights and activations as low-bit integers (int8, int16) helps to avoid the costly floating-point computation and reduces the memory footprint [3]. In this sub-section the model is quantized to 16-bit integer (int16) compared to the 32-bit floating-point models. Another benefit is that multiple data can be proceeded in single instruction with SIMD instructions. Two pairs of int16 values can be multiplied and accumulated to an int32 value in single SMLAD instruction, as shown in Figure 6. The performance of SMLAD instruction is evaluated by processing two adjacent weights and adjacent activations in one instruction. Also, the performances of unstructured pruning with skipping zero weights in normal loop order and weight prioritized order mentioned in Section 3.2 are evaluated in int16 quantized condition.

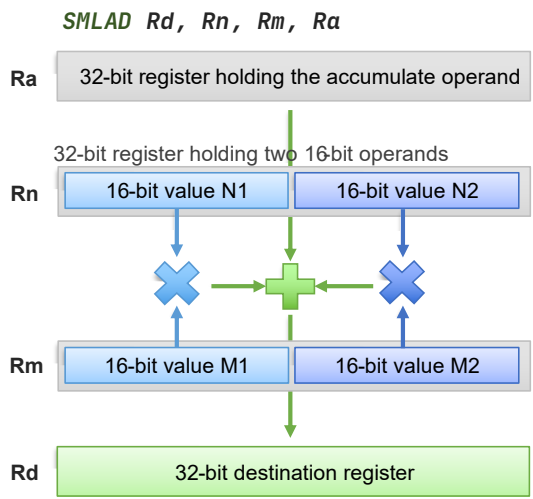

**Figure 6** Illustration for SMLAD instruction, one of SIMD instructions.

The time and power consumption for 4 different configurations are shown in Figure 7. The elapsed time is reduced from 30.8ms to 21.4ms after quantization from float32 to int16, and memory footprint reduced to almost half. Then the time is further reduced to 15.6ms when SMLAD instruction is used. Skipping zero weights under normal loop order is totally counterproductive, while skipping under weight-prioritized loop order condition still benefits when pruning percentage larger than 80%.

## 4. CONCLUSION

In this paper, we firstly showed a complete real-time small-footprint keyword spotting system running on STM32 microcontroller contains audio acquisition, feature extraction and model inference parts. It can print classification results for each ~37ms continuously. Then we evaluated the actual performance for different pruning and quantization methods on microcontroller, including different granularity of sparsity, skipping zero weights, weight-prioritized loop order, and SIMD instruction. The result shows that for microcontrollers, there are considerable challenges for accelerate unstructured pruned models, and the structured pruning is more friendly than unstructured pruning. The result also verified that the performance improvement for quantization and SIMD instruction. Further work can evaluate more acceleration methods for embedded deep learning, and effectively guide algorithms to optimize for embedded edge devices.

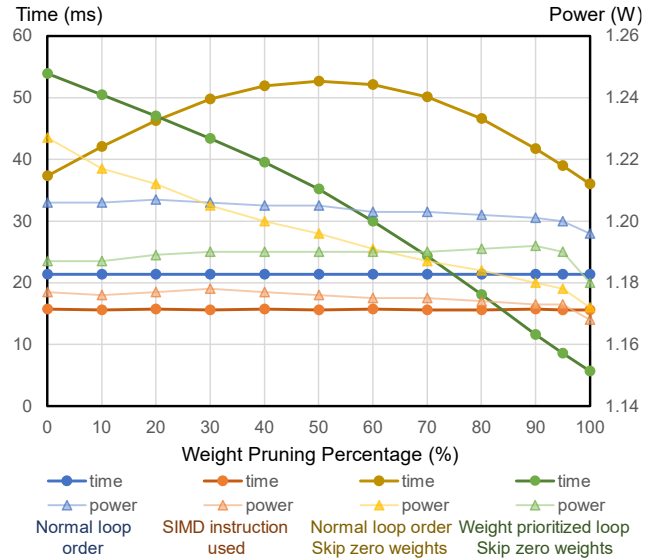

**Figure 7** Time and power consumption for int16 quantized model with/without SIMD instruction used, skipping zero weights in normal and weight prioritized loop orders.

### ACKNOWLEDGMENT

We gratitude to Chengsen Dong (Master in University of Glasgow) for our inspiring discussions about this work and the three-year-long technical exchange and friendship with the author Jingyi Wang.